# Solving Jigsaw Puzzles using Iterative Random Sampling: Parallels with Development of Skill Mastery

Neil Zhao[a,*], Diana Zheng[b]

[a]*Thomas Jefferson University, Philadelphia, Pennsylvania, USA*
[b]*St. Luke's University Health Network, Bethlehem, Pennsylvania, USA*

---

Skill mastery is a priority for success in all fields. We present a parallel between the development of skill mastery and the process of solving jigsaw puzzles. We show that iterative random sampling solves jigsaw puzzles in two phases: a lag phase that is characterized by little change and occupies the majority of the time, and a growth phase that marks rapid and imminent puzzle completion. Changes in the proportions of the number of single pieces and larger pieces can be overlaid on the timeline and progression of skill mastery. An emphasis is placed on the development of connections between pieces, which serves as an indicator of increasing puzzle completion and increasing skill mastery. Our manuscript provides a straightforward visual of skill mastery in the context of a common recreational activity.

---

*E-mail of corresponding author: neil.zhao@students.jefferson.edu

# Introduction

Jigsaw puzzles are solved by assembling a predetermined number of pieces into a unique orientation and order so as to form a single, contiguous unit that does not contain any gaps [1]. Algorithmic techniques for solving these puzzles have been developed, ranging from early methods of manipulating digitized representations of pieces into the correct order and orientation [2], analyzing boundary curvature and performing contour matching to find contiguous pieces [3,4], and prioritizing puzzle edge pieces before solving the interior [5].While these puzzles serve as recreational activities, the methods utilized to solve them can also be applied to information retrieval in piecing together shredded documents [6,7] and reconstructing faded artwork and damaged architecture [8,9].

The process of solving jigsaw puzzles also has parallels with the process of skill mastery. Malcom Gladwell introduced the concept of the 10,000 hours rule in *Outliers: The Story of Success* [10]. The rule states that 10,000 hours of mindful practice is required to master a skill. Improving cognition to achieve faster and more efficient skill mastery has also been thoroughly studied in the academic context. The increased adoption of computers and the nascent use of artificial intelligence in the classroom have been crucial components of attempts to improve student engagement and knowledge assimilation [11]. The effect of stimulants [12,13] and supplements like carotenoids [14] on academic performance has also been the subject of research into skill mastery. Various approaches to improve skill mastery in different fields have been studied for their efficacy. Simulation-based learning has seen widespread use in the medical field for resident physician education [15]. Encouraging extended periods of pause has also been studied as a potential method for improving the ability of middle school students in algebraic problem solving [16]. Similar forays into improving individual and team-based functionality using different models of teaching skill mastery have been encountered in sports [17].

While there have been extensive studies of methods for skill mastery, there is a gap of knowledge in the process by which skill mastery occurs. We present in this manuscript an algorithmic technique for jigsaw puzzle solving using iterative random sampling that can be applied to clarifying the internal dynamics of skill mastery. We show that random samplings of k pieces from a pool of unsolved jigsaw puzzle pieces eventually results in puzzle completion. This process can be divided into a lag phase and a growth phase. The lag phase is characterized by a slow increase in the size of the largest piece that is solved; while the growth phase is marked by a rapid increase in the size of the largest piece, eventually concluding in puzzle completion. We show that this change can be represented by the variation in the proportion of single and larger size pieces. Single pieces represent new, unassimilated knowledge; larger pieces represent connections that have been developed among disparate pieces of information, leading to skill mastery.

**Methods**

All analysis and simulations were performed using MATLAB R2019a (MathWorks, Natick, MA). An $n \times n$ puzzle with a total of $n^2$ pieces was solved with and without replacement (Schematic). At each iteration, k pieces were randomly sampled from the starting pool of pieces. Contiguous pieces were matched and then all pieces (single and matched) were returned to the starting pool with random sampling with replacement (Schematic A). Contiguous pieces were matched and then all sampled pieces (single and matched) were placed aside in a reserved pool until sampling size > starting pool pieces, at which point the reserved pool was returned to the starting pool (Schematic B).

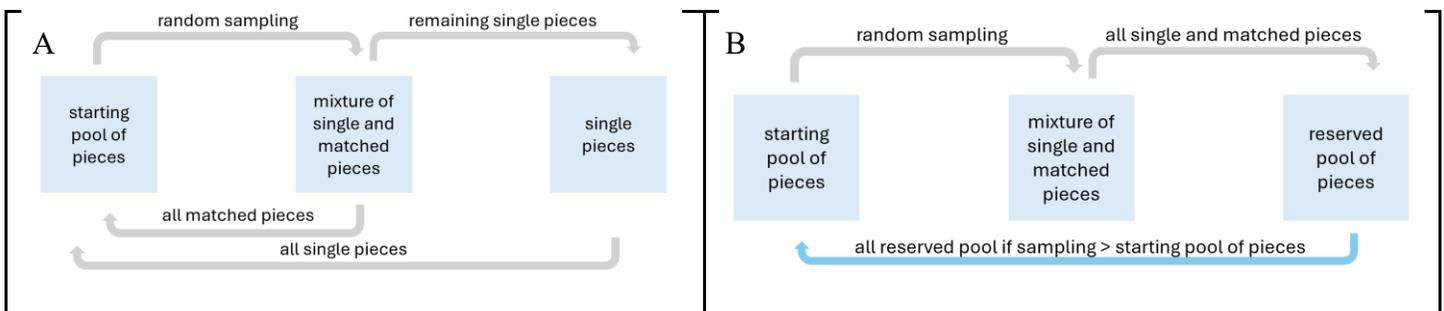

Schematic: Random sampling from a pool of puzzle pieces (A) with replacement and (B) without replacement.

Hierarchical clustering was used to ascertain contiguous pieces in each random sampling for matches. The shortest distance between points was used to establish an agglomerative hierarchical cluster tree. A cutoff of 1 in determining unique clusters based on distance was then used to distinguish between separate sets of contiguous pieces. MATLAB code provided below with annotation.

<u>Random sampling with replacement</u>

```
function [totalpieces,sizes,stepsolves,groupcolor,n]=puzzle(n,k) %n^2 is puzzle size and
%k is size of each random sample
%totalpieces tabulates the number of pieces (single+matched) after each iteration
%sizes tabulates the number of pieces and size of each piece after each iteration
%stepsolves graphs the pieces that are matched at each iteration at their locations on
the puzzle
%groupcolor encodes a unique color for all pieces that match at each iteration
a=1:n^2; %generates the n^2 pieces in puzzle, each indexed by its value in the array
sizes={};
stepsolves=[];
largest=[1]; %for tabulating the size of the largest piece at each iteration
groupcolor={};
totalpieces=[n^2];
clusters=[num2cell(a) {}]; %each unit in the starting pool holds a puzzle piece (single
or matched) and is updated at each iteration based on what is newly matched
```

```matlab
groups=[];

for i=1:1500 %for loop can be increased for larger puzzles

    if numel(clusters)==1 %will terminate function if one piece remains in puzzle,
indicating puzzle is solved
        break
    end

    grab=randperm(numel(clusters),min([k numel(clusters)]));%draws random indices (puzzle
pieces) based on cluster size
    grabsizes=cellfun(@numel,clusters(grab));%finds size of each sampled index (puzzle
piece)
    grabsizes1=grabsizes==1; %finds locations in sample of all single pieces
    grabsizes2=grabsizes>1; %finds locations in sample of all pieces size>1
    grab1=[clusters{grab}];%finds all puzzle pieces that correspond to the random
sampling

    [ally,allx]=ind2sub([n n],grab1);
    S=linkage([allx' ally'],'single');
    Y=cluster(S,'cutoff',1,'Criterion','distance'); %performs hierarchical clustering to
ascertain unique groups of contiguous pieces in each sampling
    singles=sum(Y==Y',2)==1;%finds all single pieces, without matches
    grab1(singles)=[];%removes all single pieces
    Y(singles)=[];%removes all single pieces

    clusters(grab(grabsizes2))=[]; %removes all pieces size>2 from starting pool
    pairedsizes1=[clusters{grab(grabsizes1)}];%finds indices of single sizes
    pairedIndx1=pairedsizes1(ismember(pairedsizes1,grab1)); %finds indices of single
sizes that match to form contiguous pieces in the current sampling
    pairedIndx2=cellfun(@(x2) mean(ismember(x2,pairedIndx1)),clusters); %finds the
location in the starting pool of said pieces (mean is required to covert [0 0...] to 0)
    clusters(logical(pairedIndx2))=[]; %removes all single pieces that come together to a
match from starting pool

    if ~isempty(Y)
        clustergroups=findgroups(Y);
        groupize=@(x1){x1'};
        groups=splitapply(groupize,grab1',clustergroups);
        clusters=[clusters groups']; %adds all matched pieces to the end of the starting
pool
    else
        clusters=[clusters {}];
    end

    stepsolves=cat(3,stepsolves,zeros(n));
    stepsolves(grab1+n^2*(i-1))=grab1; %records each match at each iteration
    groupcolor=vertcat(groupcolor,[{groups},{rand(numel(groups),3)}]); %assigns a unique
color to each piece that matches at each iteration

    totalpieces=[totalpieces numel(clusters)]; %tabulates number of remaining pieces at
each iteration
    clustersizes=cellfun(@numel,clusters);%finds size in each element in starting pool,
recorded at each iteration in sizes
    largest=[largest max(clustersizes)]; %records the largest sized piece at each
iteration
    sizes{i}=clustersizes;
end

figure
```
5

```matlab
scatter(1:numel(totalpieces),totalpieces,'filled') %graphs total # pieces vs iteration
ylim([0 n^2])
xlabel('Iteration')
ylabel('Total # of pieces')

figure
scatter(1:numel(largest),largest,'filled') %graphs largest piece size vs iteration
xlabel('Iteration')
ylabel('Largest piece size')
end
```

## Random sampling without replacement

```matlab
function [totalpieces,sizes,stepsolves,groupcolor,n,count]=wopuzzle(n,k)
%much of the code is the same as random sampling with replacement, will only annotate the
differences

a=1:n^2;
sizes={};
stepsolves=[];
totalpieces=[n^2];
largest=[1];
groupcolor={};
clusters=[num2cell(a) {}];
leftoverclusters={};%the reserve required to hold all sampled pieces
groups={};
count=0; %how many times the reserve is combined with the starting pool of pieces

for i=1:1500

    if numel(clusters)<k
        clusters=[clusters leftoverclusters]; %combines the reserve with the starting
pool if sampling size larger than starting pool size
        leftoverclusters={};%empties reserve pool
        count=count+1;
        if numel(clusters)==1 %terminates function if puzzle is solved
            break
        end
    end

    grab=randperm(numel(clusters),min([k numel(clusters)]));
    grabsizes=cellfun(@numel,clusters(grab));
    grabsizes1=grabsizes==1;
    grabsizes2=grabsizes>1;
    grab1=[clusters{grab}];

    [ally,allx]=ind2sub([n n],grab1);
    S=linkage([allx' ally'],'single');
    Y=cluster(S,'cutoff',1,'Criterion','distance');
    singles=sum(Y==Y',2)==1;
    singleislands=grab1(singles); %single pieces that don't match
    grab1(singles)=[];
    Y(singles)=[];

    clusters(grab)=[]; %removes all pieces from starting pool

    if ~isempty(Y)
        clustergroups=findgroups(Y);
        groupize=@(x1){x1'};
```

```matlab
            groups=splitapply(groupize,grab1',clustergroups);
            leftoverclusters=[leftoverclusters groups' num2cell(singleislands)]; %appends all
matched and single pieces that don't match to reserve pool at each iteration

        else
            leftoverclusters=[leftoverclusters num2cell(singleislands)]; %appends all single
pieces that don't match to reserve pool in case of no match pieces at each iteration
        end

        stepsolves=cat(3,stepsolves,zeros(n));
        stepsolves(grab1+n^2*(i-1))=grab1;
        groupcolor=vertcat(groupcolor,[{groups},{rand(numel(groups),3)}]);

        totalpieces=[totalpieces numel(clusters)+numel(leftoverclusters)];
        clustersizes=cellfun(@numel,[clusters leftoverclusters]);
        largest=[largest max(clustersizes)];
        sizes{i}=clustersizes;

end

figure
scatter(1:numel(totalpieces),totalpieces,'filled')
ylim([0 n^2])
xlabel('Iteration')
ylabel('Total # of pieces')

figure
scatter(1:numel(largest),largest,'filled')
xlabel('Iteration')
ylabel('Largest piece size')

end
```

**Results**

The number of remaining pieces at each iteration when solving a 10 x 10 puzzle with replacement and a sampling size of 5 pieces decreased linearly with a slope of approximately -0.5 (Fig. 1A). At iteration ~140, the magnitude of the slope increased 6X to approximately -3.0 until puzzle completion at iteration ~150. This inflection point corresponded to the size of the largest piece at each iteration. The largest size remained <5 until iteration ~80, after which the largest size increased to ~10 until iteration ~140 (Fig. 1B). Afterward, the largest size piece increased to 100 in the span of approximately 10 iterations. Two piece matches were the only solutions found with each random draw until iteration ~40, at which point the remaining pieces were composed of ~80% size 1 piece and ~20% size 2 pieces (Fig. 2A,B). The proportion of single pieces reached 60% at iteration ~100, while size 2 pieces remained at 20% and size 3 pieces were ~10%. The first appearance of a size 10 or greater piece was at iteration 80.

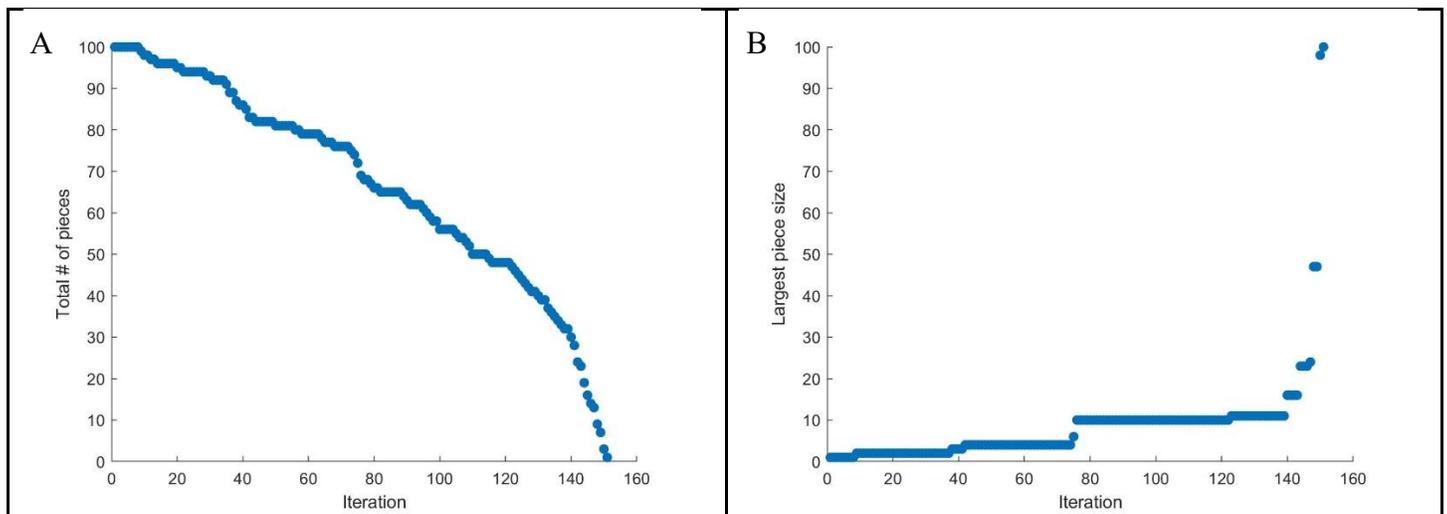

Figure 1: Iterative random draws of sampling size 5 with replacement in 10 x 10 puzzle. (A) Number of pieces and (B) size of largest piece at each iteration.

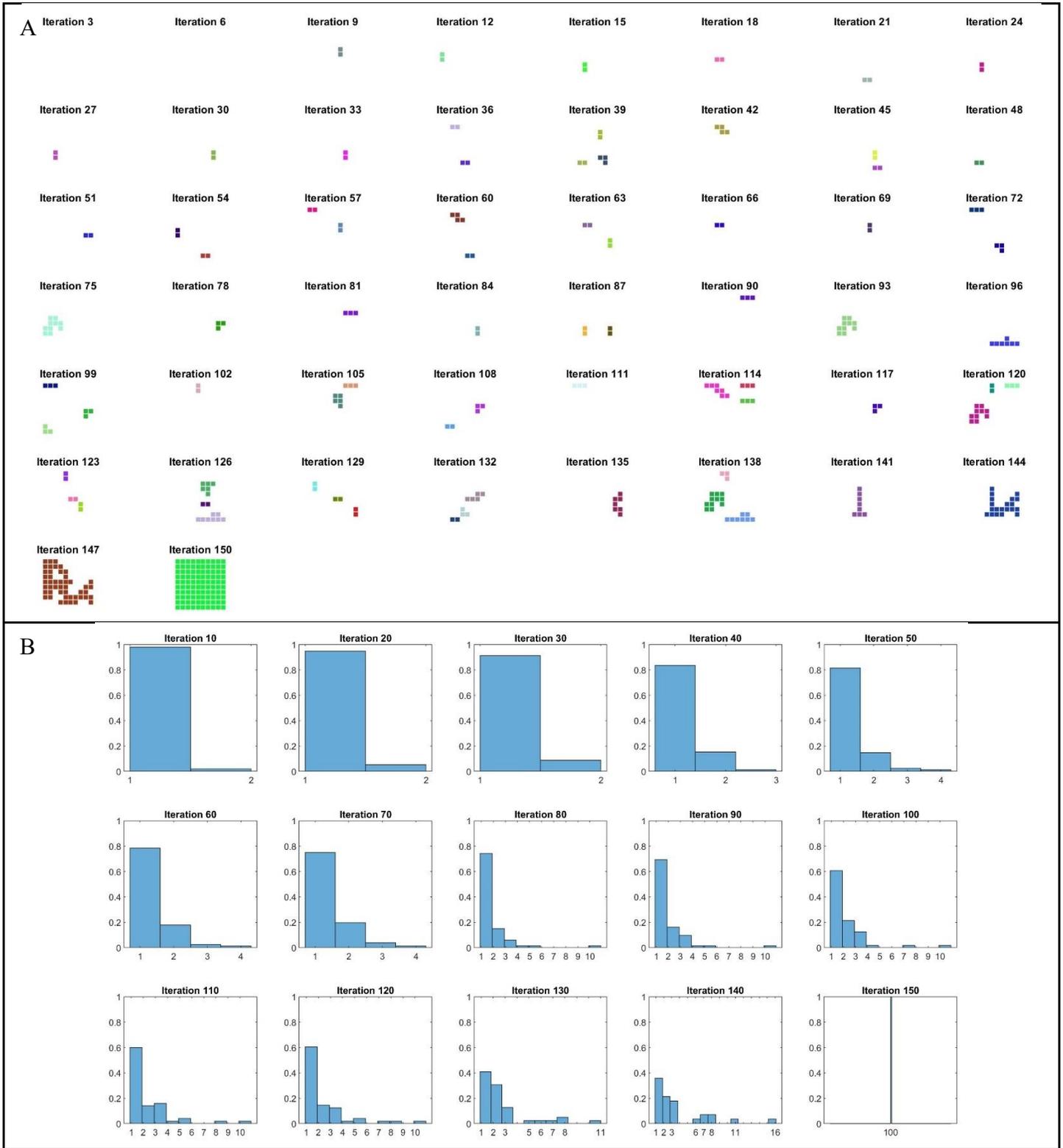

Figure 2: Iterative random draws of sampling size 5 with replacement in 10 x 10 puzzle. (A) Successful solutions and their locations on the puzzle of each draw at different iterations. (B) Percentage distribution of the sizes of all remaining pieces at different iterations.

Increasing the sampling size to 10 pieces resulted in puzzle completion with 32 iterations (Fig. 3A). A similar inflection point was encountered at iteration ~26 when the slope decreased from -2.0 to -5.5. This again corresponded with an increase in the largest piece size (Fig. 3B). Three piece matches appeared at iteration 4 with a sampling size of 10 (Fig. 4A,B). The proportion of single pieces reached 80% at iteration ~12. Size 2 pieces composed 20% and size 3 pieces composed 10% of all remaining pieces at iteration ~20. The first appearance of a size greater than 10 piece was at iteration 22.

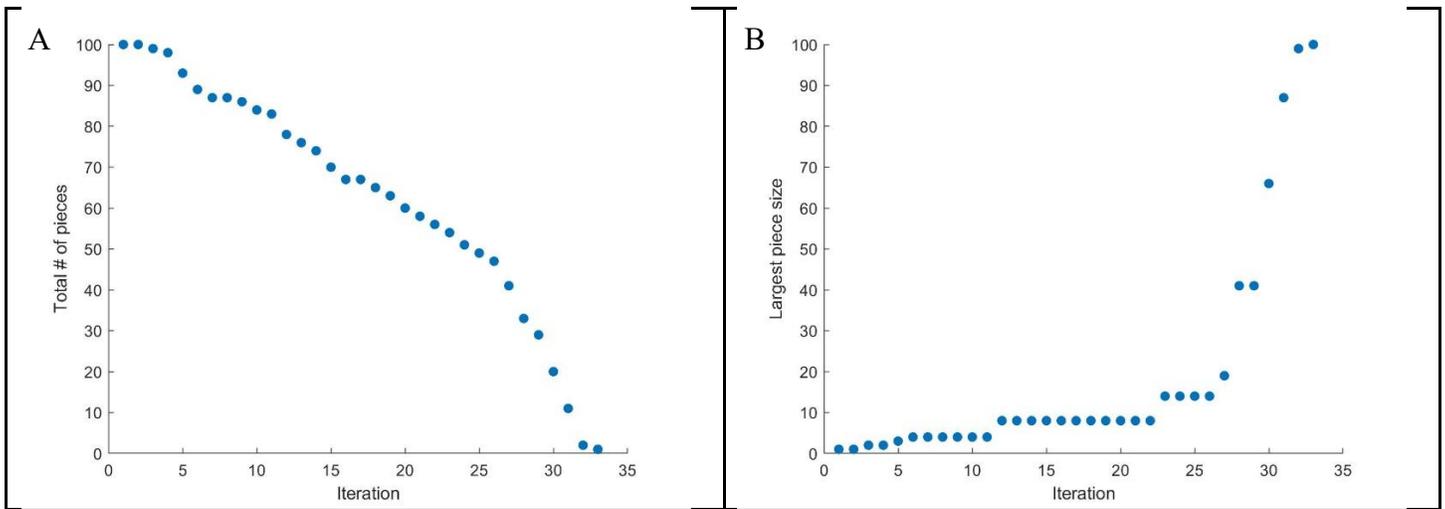

Figure 3: Iterative random draws of sampling size 10 with replacement in 10 x 10 puzzle. (A) Number of pieces and (B) size of largest piece at each iteration.

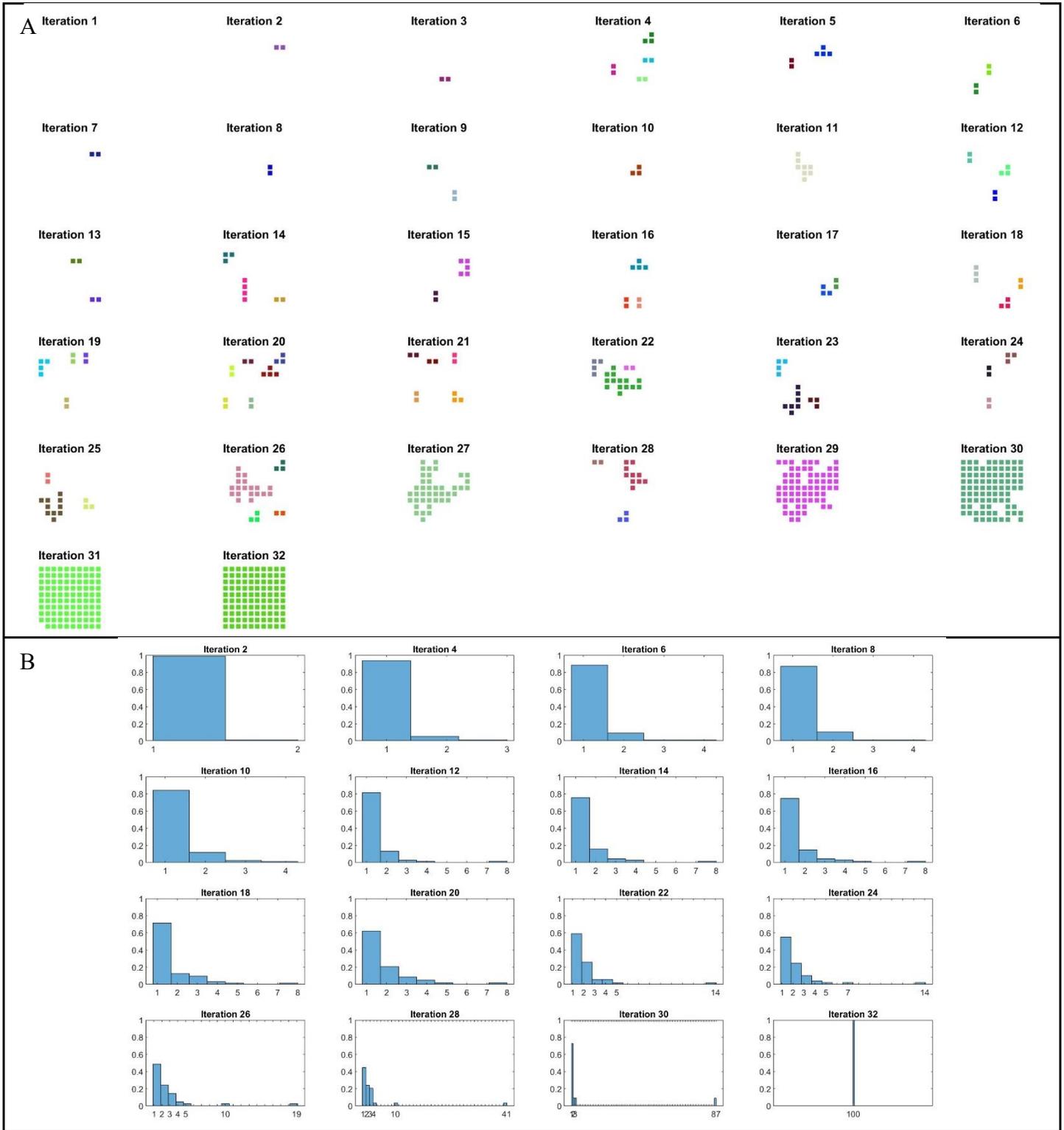

Figure 4: Iterative random draws of sampling size 10 with replacement in 10 x 10 puzzle. (A) Successful solutions and their locations on the puzzle of each draw at different iterations. (B) Percentage distribution of the sizes of all remaining pieces at different iterations.

A 20 x 20 puzzle solved with replacement and a sampling size of 10 pieces was completed in ~460 iterations (Fig. 5A). The inflection point at iteration ~400, when the magnitude of the rate of decrease of pieces increased, corresponded again to the iteration at which the size of the largest piece began to increase over shorter iteration intervals (Fig. 5B). Two piece matches remained the only solutions found with each random sampling until iteration ~90 (Fig. 6A,B). The proportion of single pieces decreased to 80% at iteration ~180, while the proportion of size 2 pieces increased to 20% at iteration ~360. The first appearance of a size greater than 10 piece was at iteration 330.

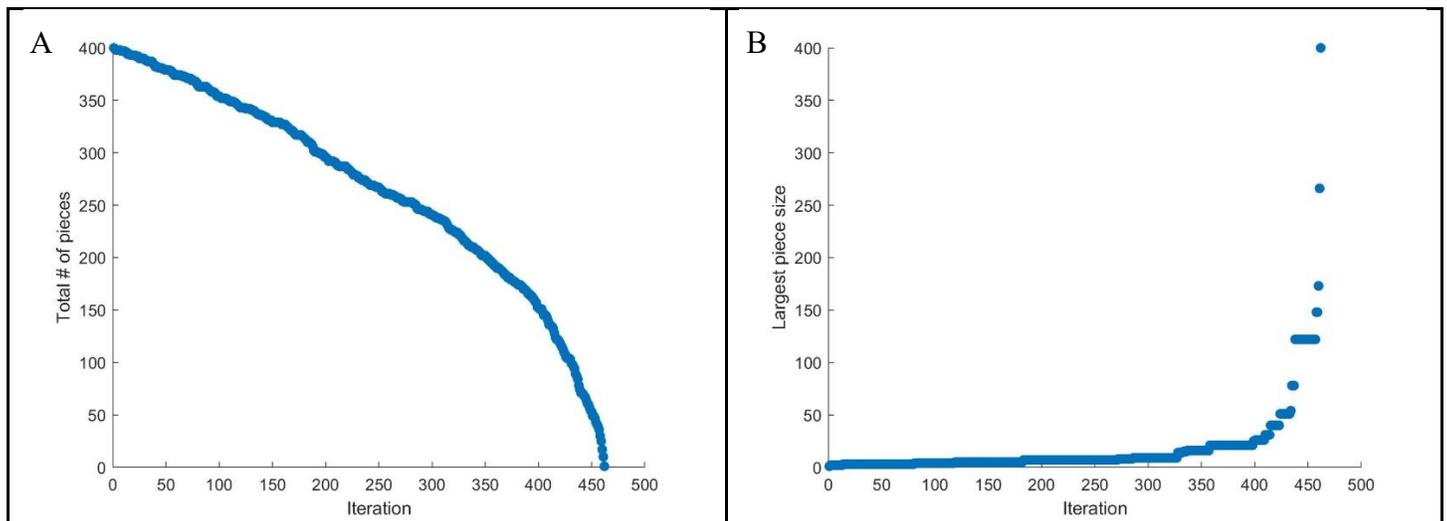

Figure 5: Iterative random draws of sampling size 10 with replacement in 20 x 20 puzzle. (A) Number of pieces and (B) size of largest piece at each iteration.

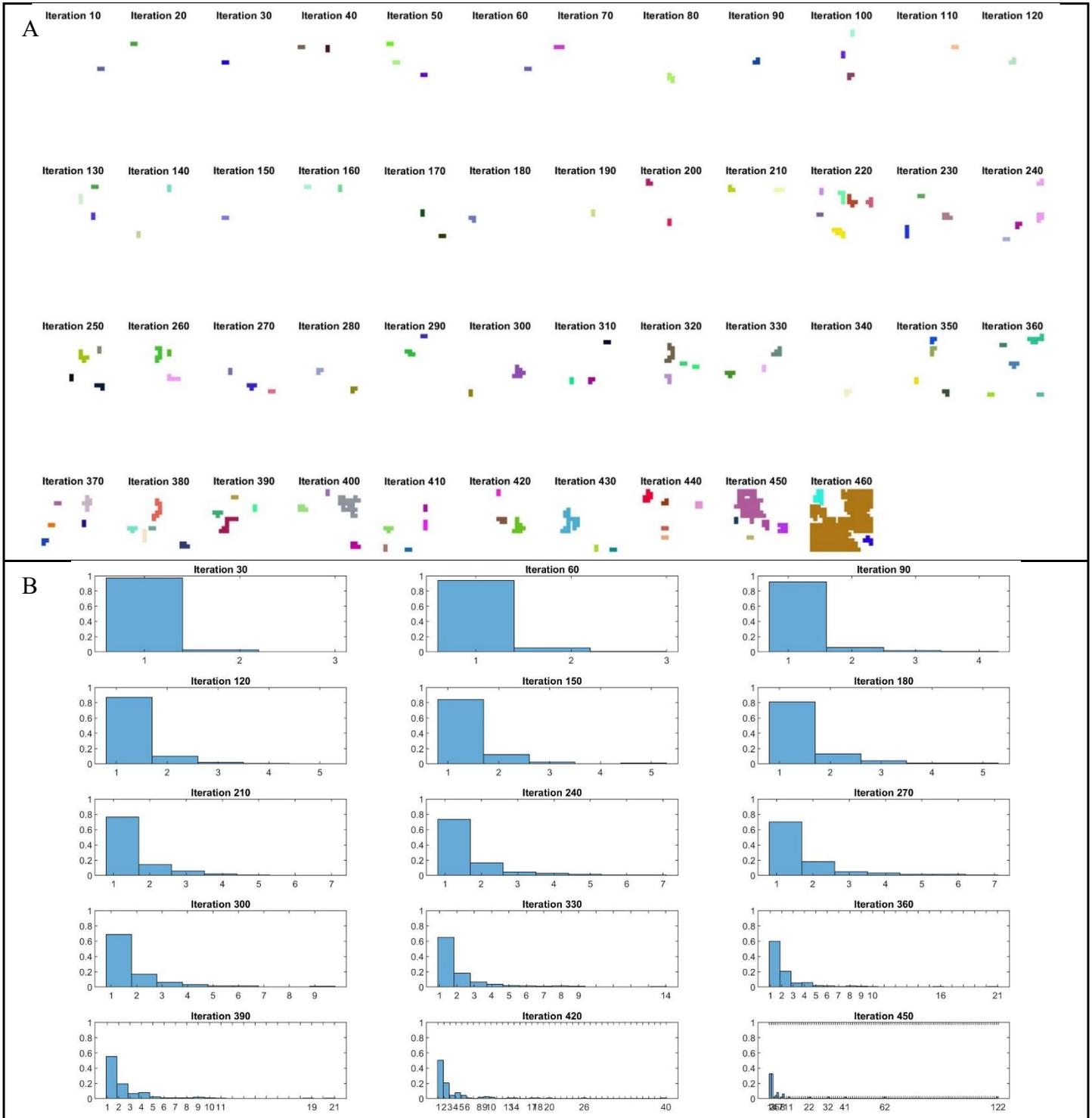

Figure 6: Iterative random draws of sampling size 10 with replacement in 20 x 20 puzzle. (A) Successful solutions and their locations on the puzzle of each draw at different iterations. (B) Percentage distribution of the sizes of all remaining pieces at different iterations.

Without replacement, a 10 x 10 puzzle with a sampling size of 5 was completed in ~170 iterations (Fig. 7A). The inflection point was not as readily identifiable from inspection, but the increase in size of the largest piece at iteration ~145 pointed to a similar location on the graph tabulating the number of pieces (Fig. 7B). Size 2 pieces solutions were uniquely present until iteration 30-40 (Fig. 8A,B). The proportion of single pieces reached 80% at iteration ~80, while the proportion of size 2 pieces reached 20% at iteration 90. Size 3 pieces reached 10% at iteration 110. The first appearance of a size greater than 10 piece was at iteration 150.

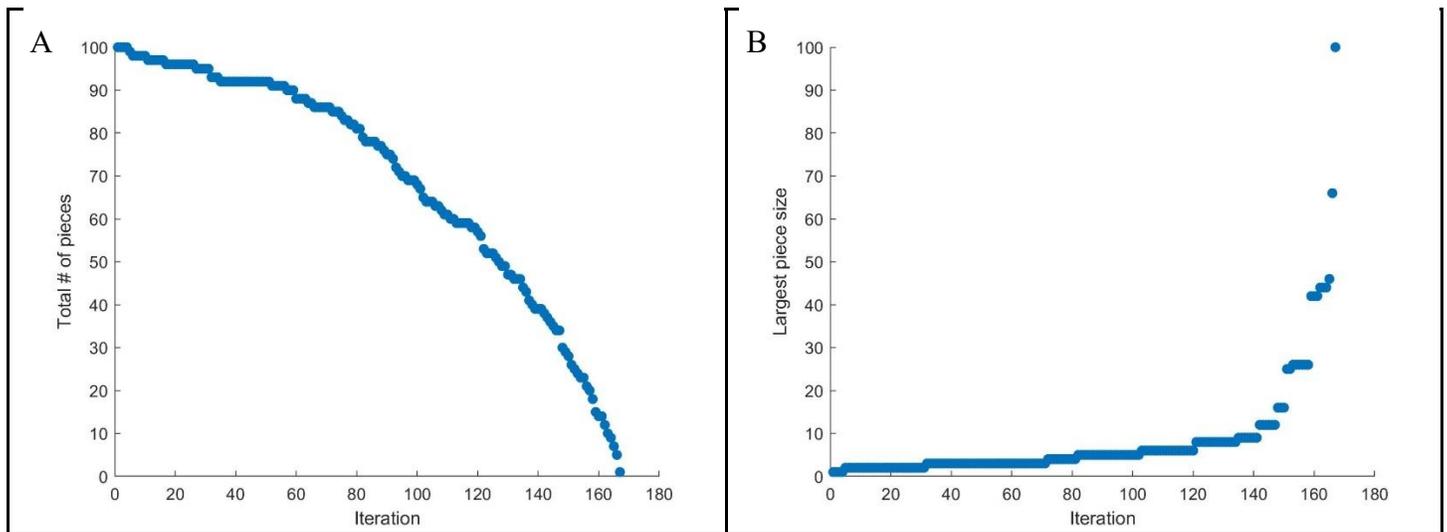

Figure 7: Iterative random draws of sampling size 5 without replacement in 10 x 10 puzzle. (A) Number of pieces and (B) size of largest piece at each iteration.

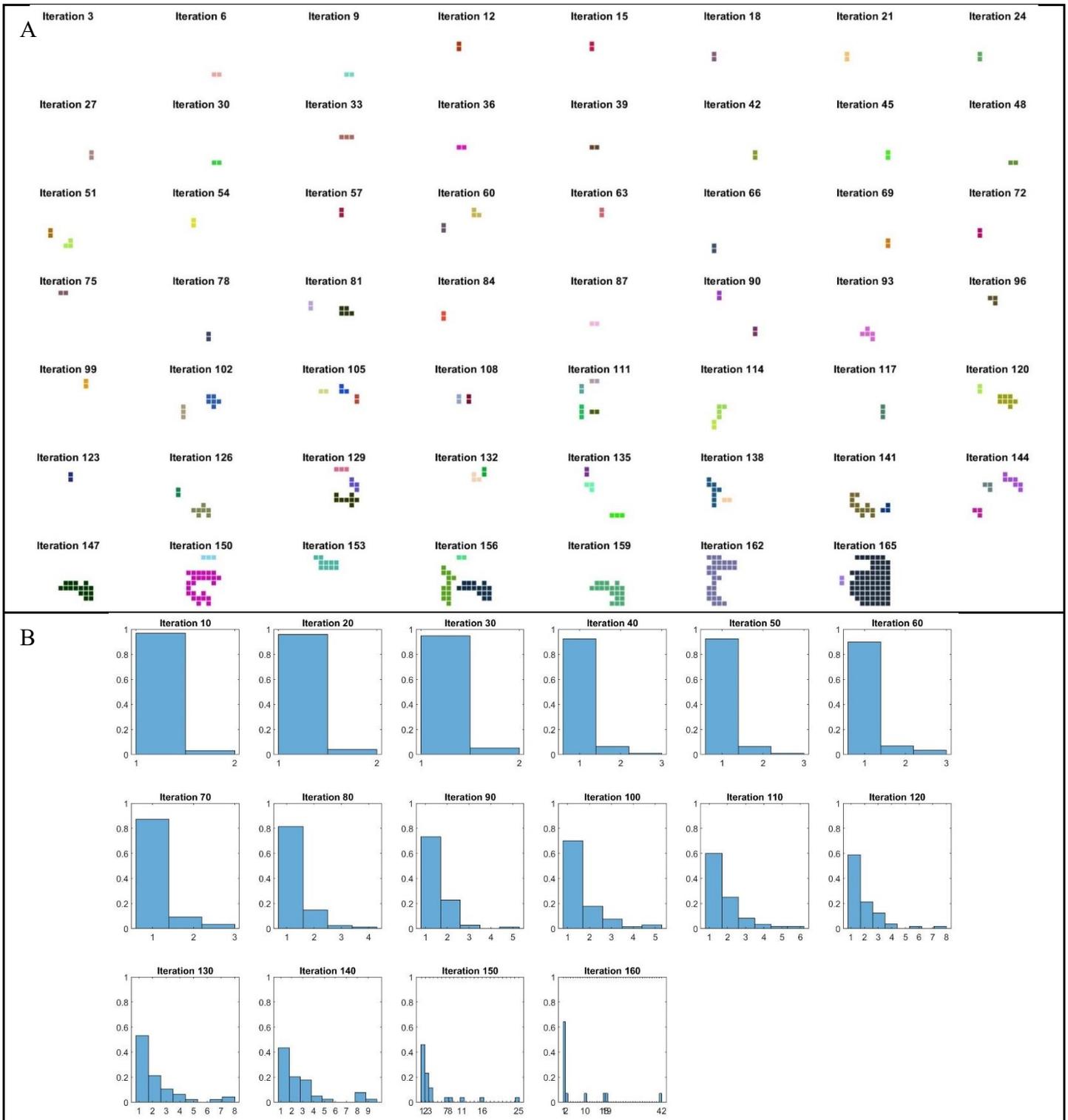

Figure 8: Iterative random draws of sampling size 5 without replacement in 10 x 10 puzzle. (A) Successful solutions and their locations on the puzzle of each draw at different iterations. (B) Percentage distribution of the sizes of all remaining pieces at different iterations.

Increasing the sampling size to 10 resulted in puzzle completion at iteration 35 (Fig. 9A). In this case, an inflection point in slope appeared at iteration ~30, which was also reflected in the change in size of the largest piece at the same iteration (Fig. 9B). The proportion of single pieces decreased to 80% at iteration 14; and size 3 pieces appeared at iteration 2 (Fig. 10A,B).

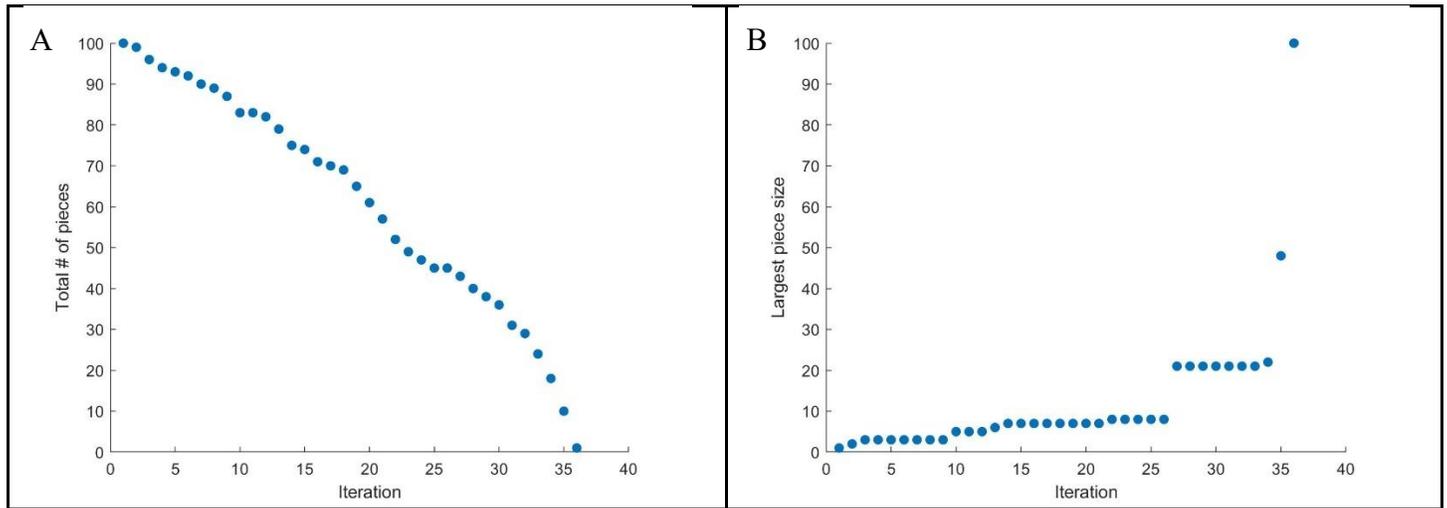

Figure 9: Iterative random draws of sampling size 10 without replacement in 10 x 10 puzzle. (A) Number of pieces and (B) size of largest piece at each iteration.

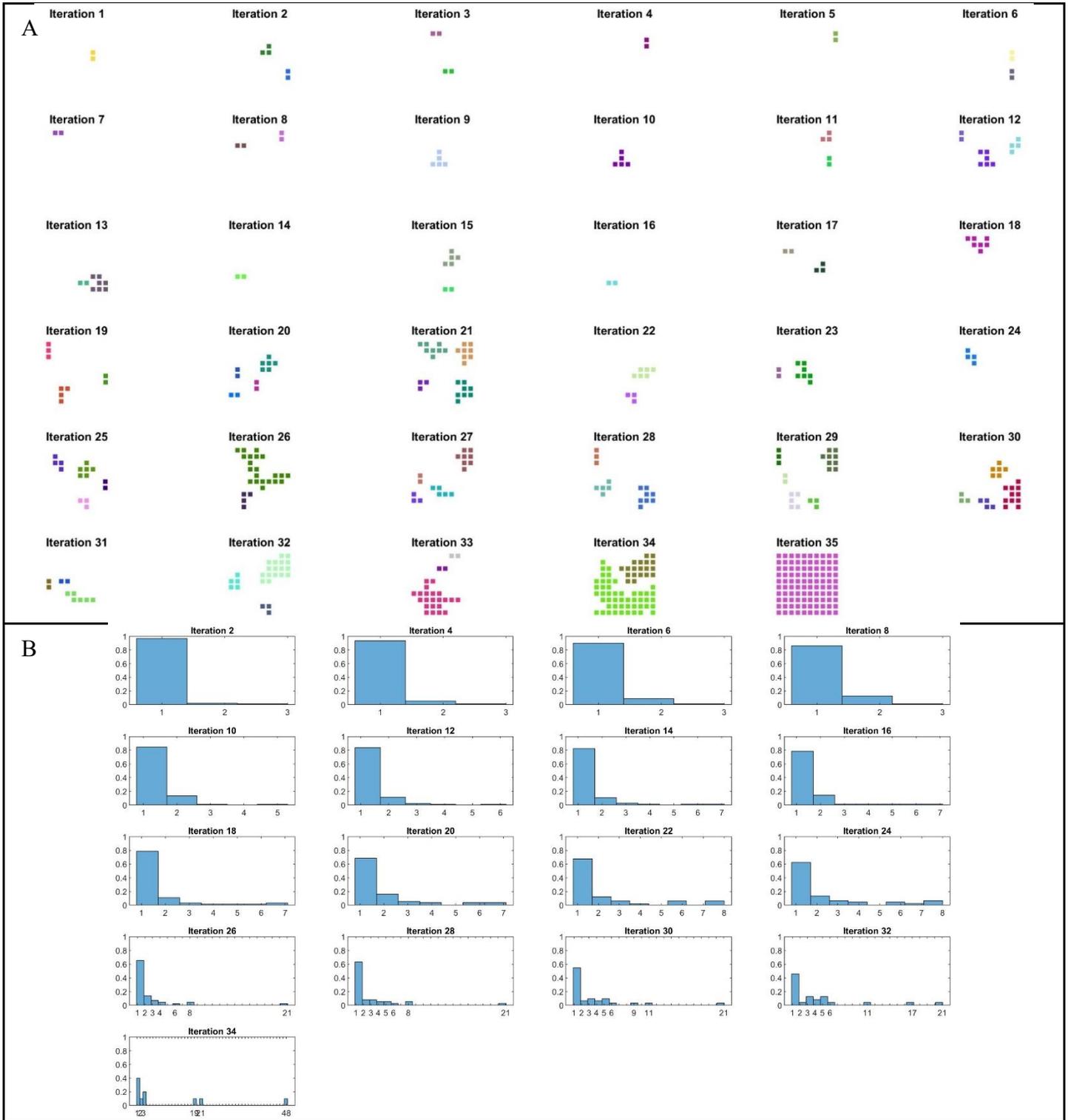

Figure 10: Iterative random draws of sampling size 10 without replacement in 10 x 10 puzzle. (A) Successful solutions and their locations on the puzzle of each draw at different iterations. (B) Percentage distribution of the sizes of all remaining pieces at different iterations.

A 20 x 20 puzzle was also solved without replacement and with a sampling size of 10. The puzzle was completed in ~450 iterations (Fig. 11A,B). The size of the largest piece began to increase nonlinearly at iteration

~375, which corresponded to an inflection point in which the magnitude of the rate of change of pieces increased.

The proportion of single pieces reached 80% at iteration 150, and the proportion of size 2 pieces reached 20% at iteration 300 (Fig. 12A,B). The first appearance of a size greater than 10 piece was at iteration 240.

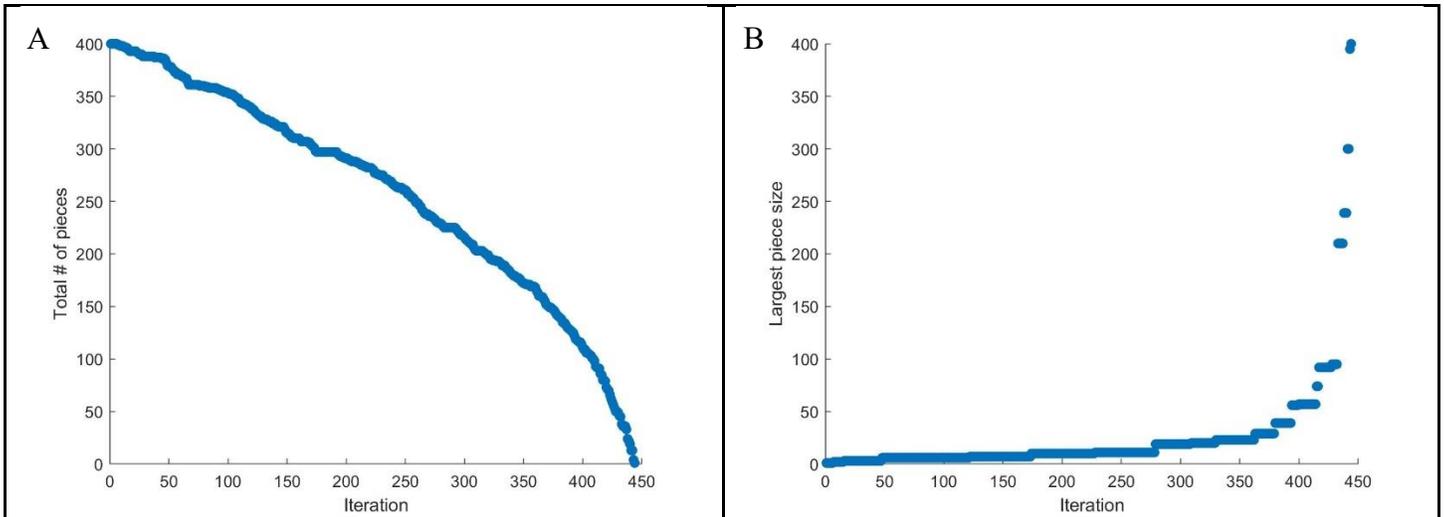

Figure 11: Iterative random draws of sampling size 10 without replacement in 20 x 20 puzzle. (A) Number of pieces and (B) size of largest piece at each iteration.

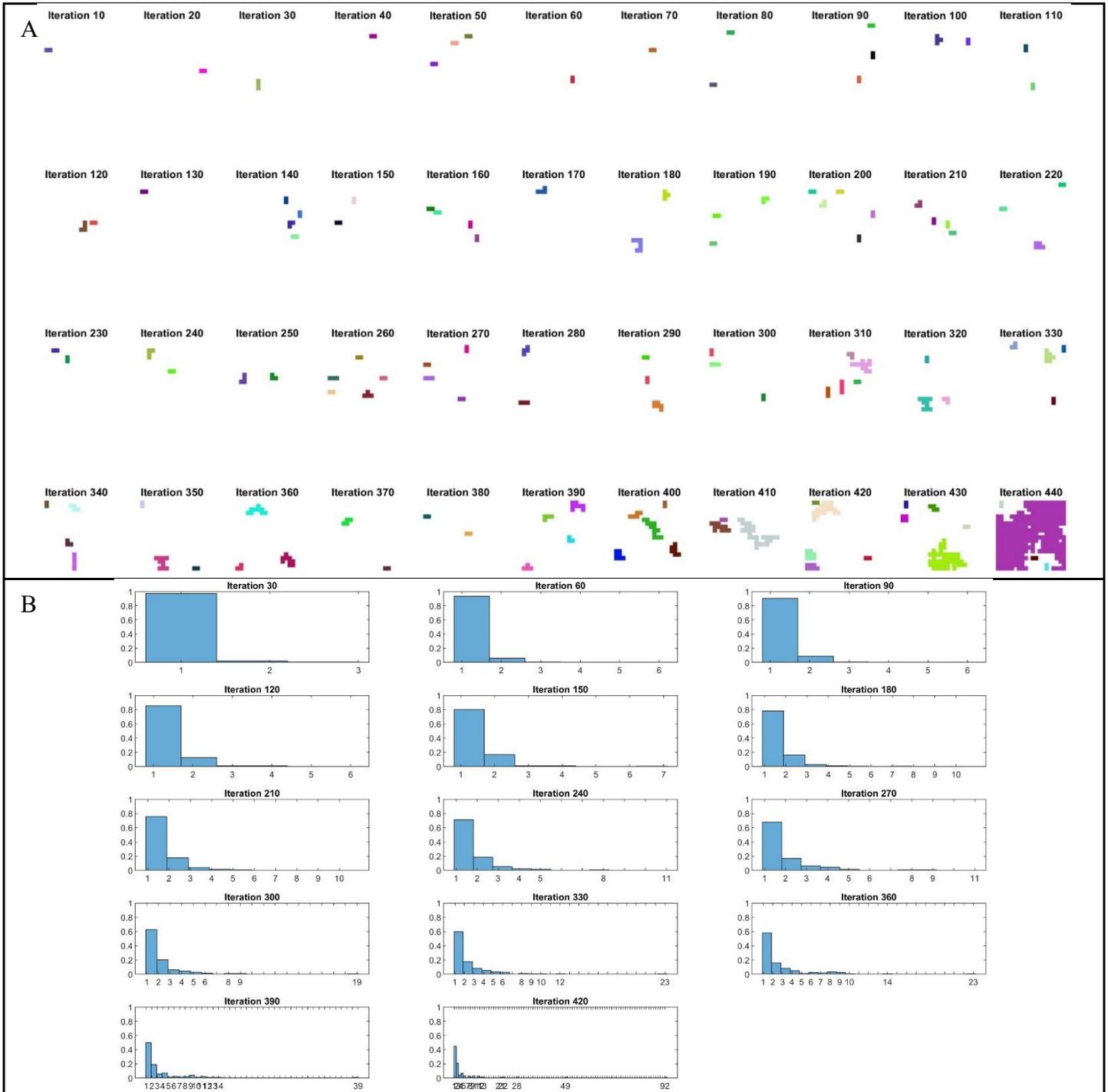

Figure 12: Iterative random draws of sampling size 20 without replacement in 20 x 20 puzzle. (A) Successful solutions and their locations on the puzzle of each draw at different iterations. (B) Percentage distribution of the sizes of all remaining pieces at different iterations.

A 10 x 10 puzzle solved with replacement and a sampling size of 2 most accurately simulates the conditions of a realistic attempt at solving a jigsaw puzzle. The puzzle was solved in ~1450 iterations; and as before the magnitude of the rate of change of puzzle pieces increased at iteration ~1250, which corresponded to a similar change in magnitude of the slope in the size of the largest piece at the same iteration (Fig. 13A,B). Size 2 pieces were the only solutions encountered in the random samplings until iteration ~450 (Fig. 14A,B). The proportion of single pieces decreased to 80% at iteration 700, while the proportion of size 2 pieces increased to 20% at iteration 850. The first appearance of a size greater than 10 piece was at iteration 1200.

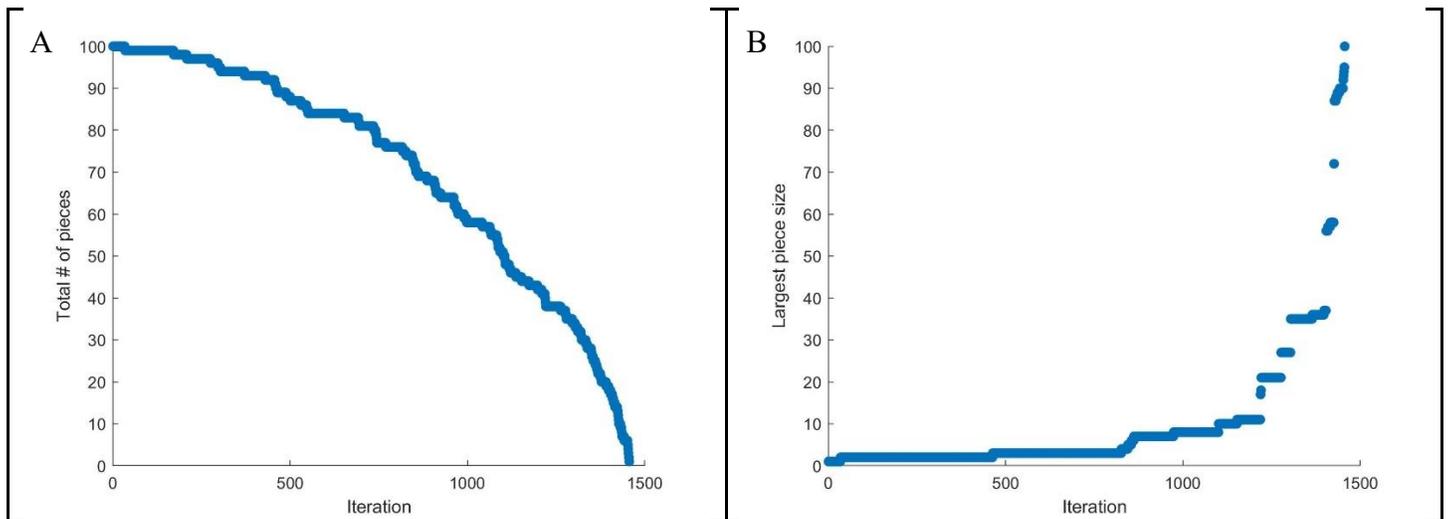

Figure 13: Iterative random draws of sampling size 2 with replacement in 10 x 10 puzzle. (A) Number of pieces and (B) size of largest piece at each iteration.

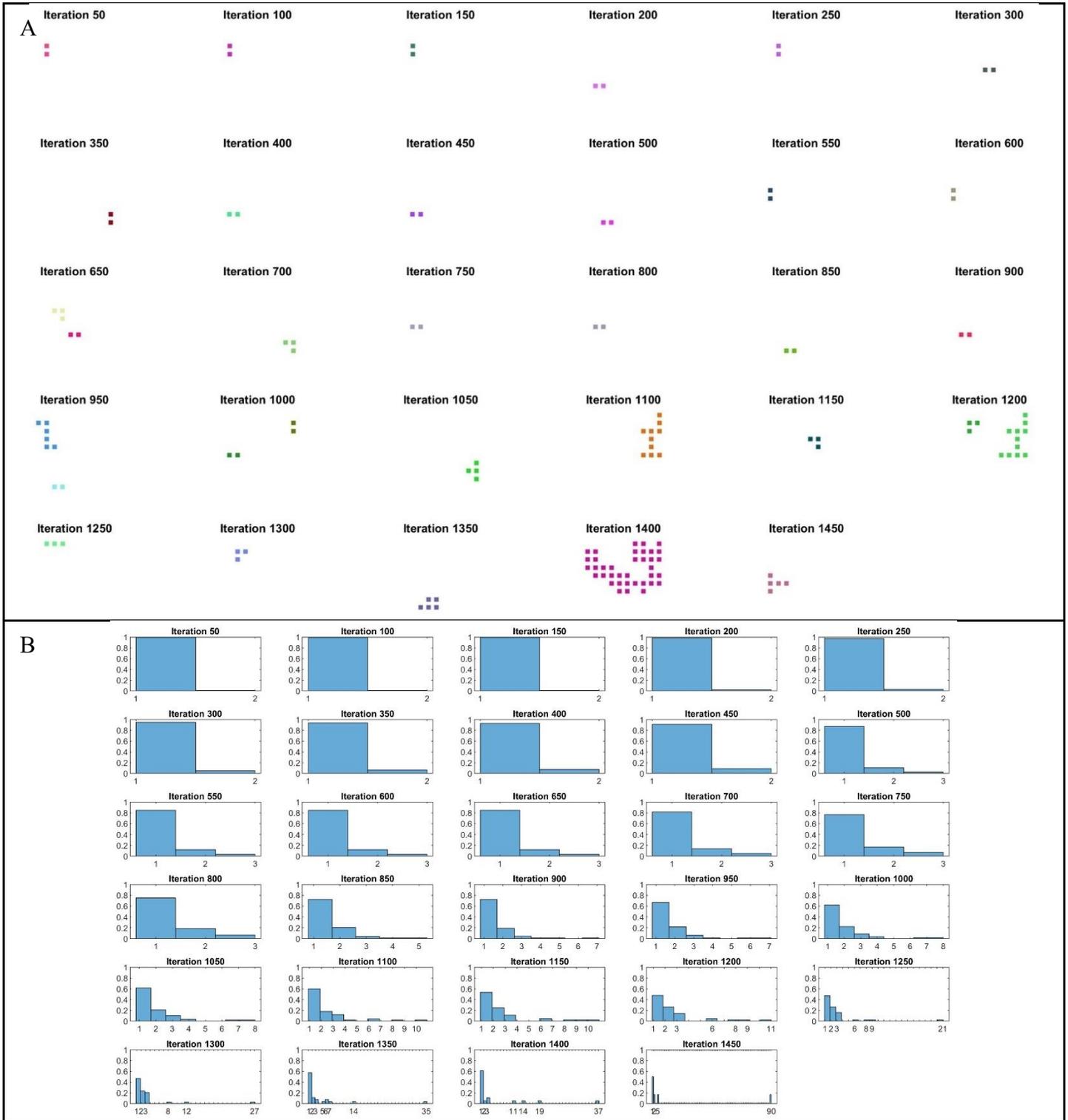

Figure 14: Iterative random draws of sampling size 2 with replacement in 10 x 10 puzzle. (A) Successful solutions and their locations on the puzzle of each draw at different iterations. (B) Percentage distribution of the sizes of all remaining pieces at different iterations.

The number of iterations with replacement required to reduce the number of pieces in a 10 x 10 puzzle by half decreased with sampling size (Table 1). An average of 972.45 iterations were required to achieve half completion with a sampling size of 2. This decreased by ~6.0X to 160.10 iterations with a sampling size of 4. Increase the sampling size to 6 resulted in half completion at an average of 65.85 iterations, which was only a decrease of ~2.5X with respect to sample size 4. Further increase to a sampling size of 8 resulted in puzzle half completion at iteration 36.60, or ~1.8X decrease compared to sampling size 6. A sampling size of 2 was able to achieve full completion at an average of 1313.40 iterations; this decreased to 218.40 iterations with a sampling size of 4 (Fig. 15B) and 92.30 with a sampling size of 6. At the high end, increasing the sampling size from 18 to 20 decreased the average iteration for half completion from 7.80 to 6.70 and decreased that for full completion from 11.95 to 10.35. This nonlinear dependence between sampling size and iteration to half or full completion can be visualized in (Fig. 15A,B).

The largest change in iterations until half or full completion of a 10 x 10 puzzle with replacement was encountered at sampling sizes 2 to 4. The proportion of single pieces decreased linearly until iteration 600 with sampling size 2 (Fig. 15C). There was a corresponding linear increase in the proportion of size 2-10 pieces. Size 11-50 pieces appeared at iteration ~900; and size 51-100 pieces appeared at iteration ~1100. Increasing the sampling size to 4 resulted in a similar linear change in the proportion of single and size 2-10 pieces (Fig. 15D). However, size 11-50 pieces appeared at iteration ~170 and size 51-100 pieces appeared at iteration ~225.

| Iteration at half completion | | | Iteration at full completion | | |
|---|---|---|---|---|---|
| k | Average | Standard deviation | k | Average | Standard deviation |
| 2 | 972.45 | 128.23 | 2 | 1313.40 | 113.73 |
| 4 | 160.10 | 26.84 | 4 | 218.40 | 22.94 |
| 6 | 65.85 | 9.03 | 6 | 92.30 | 9.09 |
| 8 | 36.80 | 5.04 | 8 | 50.45 | 4.63 |
| 10 | 23.60 | 3.28 | 10 | 33.40 | 3.72 |
| 12 | 16.80 | 1.61 | 12 | 24.40 | 1.85 |
| 14 | 11.55 | 1.10 | 14 | 17.35 | 1.39 |
| 16 | 10.00 | 1.08 | 16 | 14.85 | 1.18 |
| 18 | 7.80 | 0.77 | 18 | 11.95 | 0.83 |
| 20 | 6.70 | 0.66 | 20 | 10.35 | 0.67 |

Table 1: Iterative random draws of different sampling sizes with replacement in 10 x 10 puzzle to achieve half and full completion.

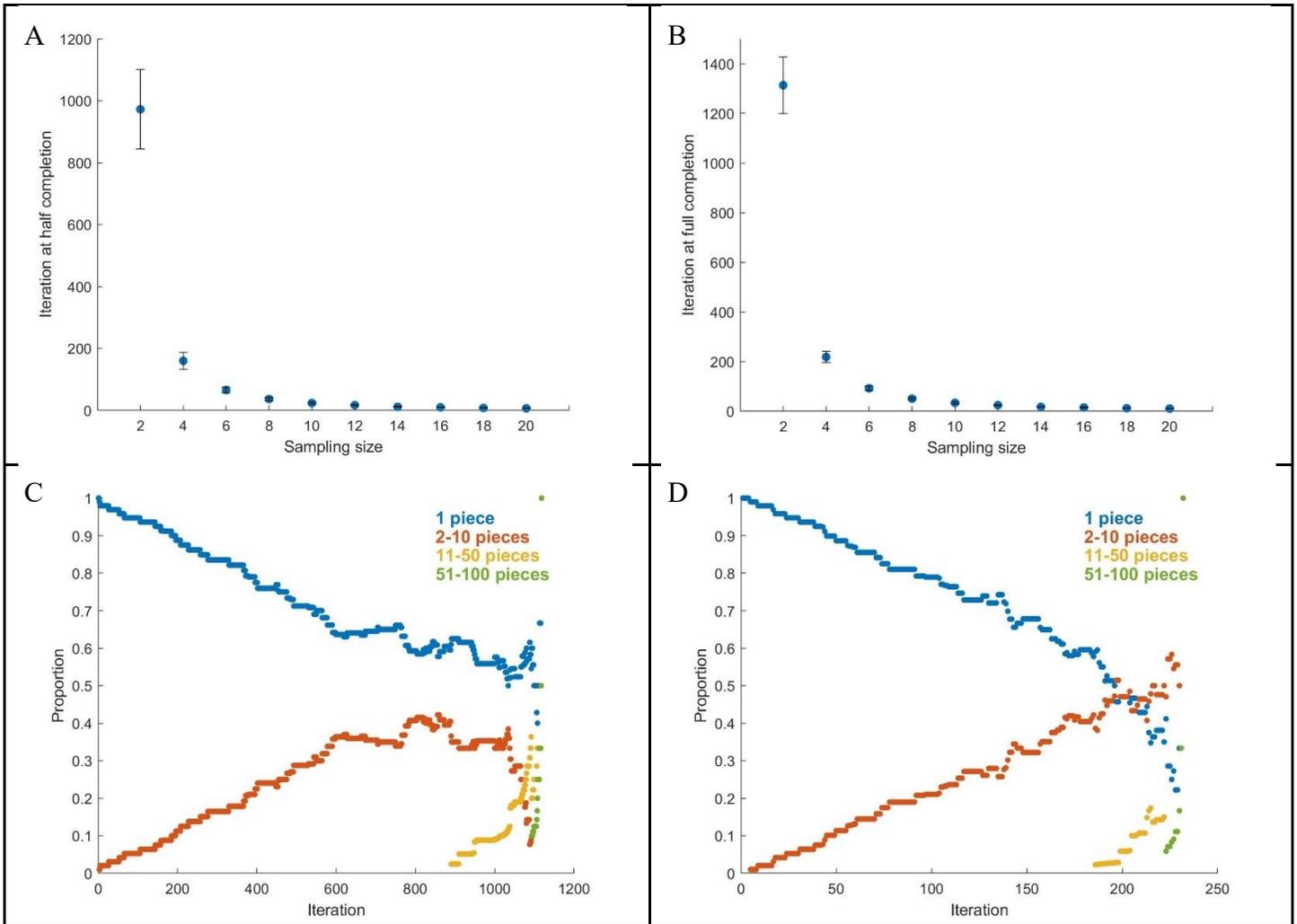

Figure 15: Iterative random draws of different sampling sizes with replacement in 10 x 10 puzzle. Iterations to achieve (A) half completion and (B) full completion. Plotted are means ± standard deviations; n=20 for each data point. Proportion of differently sized pieces at each iteration with sampling sizes (C) 2 and (D) 4.

## Discussion

A typical approach in solving jigsaw puzzles begins with completing the border. A combination of random sampling and pattern recognition is then implemented to solve the interior. Iterative random sampling is an approach to solving jigsaw puzzles that does not adhere to any sequence of priorities or relies upon real-time pattern recognition. In fact, the typical approach to solving the interior of a jigsaw puzzle can be modeled with iterative random samplings of size 2. We have shown that many iterations are required in this approach, both with and without replacement, to achieve puzzle completion. Furthermore, the number of iterations has a nonlinear, negative dependence on sampling size. An increase in sampling size from 2 to 4 results in a larger decrease in the number of required iterations for puzzle completion than encountered when increasing the sampling size from 18 to 20. Parallel with this is the observation that the size of the largest piece with each iteration can be separated into a lag phase and a growth phase. During the lag phase, the size of the largest piece remains roughly the same with each iteration. During the growth phase the size of the largest piece drastically increases over the span of a few iterations, quickly reaching puzzle completion. Additionally, the lag phase occupies the majority of the iterations, while the growth phase only appears closer to the end. These two phenomena are unified in our analysis of puzzle piece size variation with each iteration. The total number of single pieces decreases linearly, while the total number of size 2-10 pieces increases linearly with each iteration. At a timepoint closer to puzzle completion, larger pieces of sizes 11-50 or 51-100 emerge and rapidly increase. A notable observation is that the emergence of the larger sized pieces is around iteration 900 for a sampling size of 2 but is around iteration 170 for a sampling size of 4.

Parallels can be drawn between our observations of puzzle piece size emergence and the process of skill mastery in different fields. Initial entrance into a new field is frequently accompanied by the difficulty of learning new skills. Mastery of these skills derives from learning and connecting new information relevant to these new fields. A common occurrence is that the availability of new information is so large as to make connections between them difficult to form. This process of exposure to new input can be modeled as a random sampling. During the early iterations, each random sampling will be unlikely to result in substantial connections due to the large pool of information. However, as the number of iterations increases, the likelihood of a large connection emerging increases. The emergence of these large connections is also a marker of imminent puzzle completion, or skill mastery. Increasing the sampling size is another way to increase the likelihood of formation connections between different datapoints. New information can be viewed as size 1 pieces; elementary mastery can be viewed as size 2-10 pieces; advanced mastery can be viewed as size 11-50 pieces; and complete mastery can be viewed as size 51-100 pieces.

Such a model for skill mastery is based on iterative random sampling for solving jigsaw puzzles. One may conclude that the speed of mastery correlates positively with 1) the sampling size, or the amount of information one chooses to acquire and 2) the rate at which new iterations occur, or the rate one repeats the sampling.